\documentclass[conference]{IEEEtran}
\IEEEoverridecommandlockouts
\usepackage{cite}
\usepackage{balance}
\usepackage{soul}
\usepackage{amsmath,amssymb,amsfonts,amsthm}
\usepackage{algorithmic}
\usepackage{graphicx}
\usepackage{textcomp}
\usepackage{xcolor}
\usepackage{tikz}
\def\BibTeX{{\rm B\kern-.05em{\sc i\kern-.025em b}\kern-.08em
    T\kern-.1667em\lower.7ex\hbox{E}\kern-.125emX}}

\usepackage{comment}
\theoremstyle{plain}
\newtheorem*{problem}{Problem}
    
\begin{document}

\title{On the Potential of an Independent Avatar to Augment Metaverse Social Networks}

\author{
\IEEEauthorblockN{Theofanis P. Raptis, Chiara Boldrini, Marco Conti, and Andrea Passarella}
\IEEEauthorblockA{Institute for Informatics and Telematics, National Research Council, Italy\\
Email: \small \{theofanis.raptis, chiara.boldrini, marco.conti, andrea.passarella\}@iit.cnr.it
}}

\maketitle
\begin{tikzpicture}[remember picture,overlay]
\node[anchor=south,yshift=10pt] at (current page.south) {\fbox{\parbox{\dimexpr\textwidth-\fboxsep-\fboxrule\relax}{
  \footnotesize{
     \copyright 2024 IEEE. Personal use of this material is permitted.  Permission from IEEE must be obtained for all other uses, in any current or future media, including reprinting/republishing this material for advertising or promotional purposes, creating new collective works, for resale or redistribution to servers or lists, or reuse of any copyrighted component of this work in other works.
  }
}}};
\end{tikzpicture}

\begin{abstract}
We present a computational modelling approach which targets capturing the specifics on how to virtually augment a Metaverse user's available social time capacity via using an independent and autonomous version of her digital representation in the Metaverse. We motivate why this is a fundamental building block to model large-scale social networks in the Metaverse, and emerging properties herein. We envision a Metaverse-focused extension of the traditional avatar concept: An avatar can be as well programmed to operate independently when its user is not controlling it directly, thus turning it into an agent-based digital human representation. This way, we highlight how such an independent avatar could help its user to better navigate their social relationships and optimize their socializing time in the Metaverse by (partly) offloading some interactions to the avatar. We model the setting and identify the characteristic variables by using selected concepts from social sciences: ego networks, social presence, and social cues. Then, we formulate the problem of maximizing the user's non-avatar-mediated spare time as a linear optimization. Finally, we analyze the feasible region of the problem and we present some initial insights on the spare time that can be achieved for different parameter values of the avatar-mediated interactions.
\end{abstract}


\section{Introduction}

The Metaverse refers to a ``created world’’ \cite{HWANG2022100082} that has been discussed in various forms over the years, but it has recently gained a lot of attention due to its potential to revolutionize the way we interact with other people and entities through virtual and augmented environments. In essence, the Metaverse is a fully or partially virtual world that is entirely interconnected, and it consists of a complex network of physical, virtual, and social elements. The Metaverse can be thought of as an evolution of the Internet and social media, but with a more immersive and interactive experience \cite{10376184}. 

The emergence of the Metaverse is introducing a new era of cyber-physical interconnectedness, offering users unprecedented opportunities for social interaction and engagement, pushing the limit of the Web as we know it today \cite{10368182}. In this promising landscape, the concept of avatars and agents has been a focal point~\cite{9944868}. \emph{Avatars} 
are digital representations or embodiments of users within a virtual environment, often taking the form of graphical or 3D models. These representations are controlled by users to navigate and interact with the digital space, reflecting the user's identity and actions. \emph{Agents} 
are digital entities that operate autonomously, representing computer algorithms or scripted behaviors. Unlike avatars, agents can perform actions and make decisions independently of direct user input, exhibiting a certain level of intelligence and functionality. The field of social sciences is, in fact, exploring the integration of such virtual entities, to understand and enhance human experiences in the Metaverse.

\begin{figure}[b!]
    \centering
    \includegraphics[width=\columnwidth]{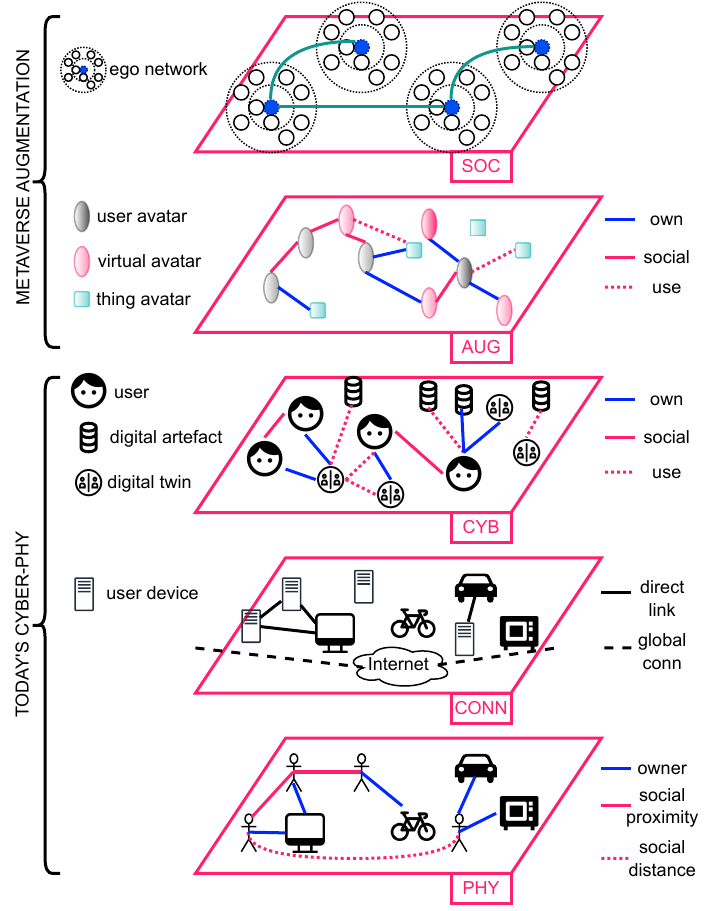}
    \caption{Logical layers of the Metaverse.}
   \label{fig::logical}
\end{figure}

Through this prism, we can view the Metaverse as not just an extension of the current networked cyber-physical setting \cite{10239355}, but a new paradigm altogether. Specifically, in our view, as depicted in Figure \ref{fig::logical}, the Metaverse consists of two entirely new layers, the augmentation layer (AUG) and social layer (SOC), which can be materialized on top of the existing physical, connectivity, and cyber layers of the traditional cyber-physical setting. The augmentation layer represents the (fully or partially) virtual realm which is used to create a sense of presence and immersion for the users, and the social layer represents the augmented interactions between the human users, virtual counterparts and connected objects. Together, these layers can potentially fuse a highly immersive and interactive environment that offers a wider spectrum of capabilities for exploration, socialization, collaboration and innovation. Specifically, at the social layer, novel forms of social networking can emerge, whereby interactions are a mix of current social networking, interaction between real users and avatars of different kinds, and direct interactions between avatars/agents. These two new logical layers may drastically modify social networking services as we know them today. In this work we wish to investigate the new form of social networking enabled by the Metaverse.

The social networking aspects of the Metaverse, and especially when it comes to its computational modelling, are in an embryonic stage in the literature. Some studies focus on the virtual reality requirements and capabilities, but there has been no work addressing the modelling of the effect of the novel Metaverse concepts (e.g., the presence of avatars/agents) on the characteristics of users' social networks. 
For example, in \cite{ARPACI2022102120}, the authors explore the social sustainability of the Metaverse by developing a model on the big five personality traits to understand the sociality aspects of the Metaverse. Or, in \cite{OH2023107498}, the authors investigate the social benefits of using Metaverse platforms, and whether enhanced social presence in the Metaverse facilitates supportive interactions among young users. 
In \cite{10.1145/3511808.3557487}, the authors explore a scenario of socialising in Metaverse with virtual reality, by formulating the co-presence and occlusion-aware Metaverse user recommendation problem. In their case they focus more on the interplay of hardware technological enablers with social presence concepts. 
In \cite{10.1145/3517745.3561417}, the authors systematically study the network performance of social activities in five popular social virtual reality platforms. Their experimental results reveal that all these platforms are still in their early stage and face fundamental technical challenges to realise the grand vision of social Metaverse. Specifically, they identify the platform servers' direct forwarding of avatar data for embodying users without further processing as the main reason for the poor scalability and discuss potential solutions to address this problem.
In our paper, we instead focus on the never-researched-before potential of independent avatars to augment the user social networking time capacity, and start shedding light on the potential effects on their social network structure, as a key element to characterise the properties of future large-scale Metaverse Online Social Networks.

Traditionally, avatars have been perceived as human-controlled entities, extensions of users' digital personas actively manipulated in real-time during interaction \cite{Sibilla_Mancini_2018}. However, in this paper, we focus on a Metaverse-centered extension of the avatar definition where avatars transcend their conventional roles, empowered to operate autonomously, akin to agents, when not under direct user control; we dub them \emph{independent avatars}. This evolution effectively transforms avatars into agent-based digital human representations, capable of independently navigating the digital social landscape and performing standalone socialization tasks.

The motivation behind this approach lies in the optimization of users' social time within the Metaverse \cite{MOGAJI2023102659}. Specifically, it has been shown that user socialisation time determines the structure of \emph{ego networks} (better described in Section~\ref{sec:vision}), i.e., the way an individual (ego) organises social direct relationships with their peers (alters). In turn, the properties of ego networks have been shown to determine macroscopic properties at the network level (e.g., information diffusion)~\cite{DBLP:journals/osnm/ArnaboldiCPD17}. Moreover, models have been proposed to build large-scale social networks starting from microscopic models of ego networks~\cite{Conti2012}, as the basis for studying unexpected phenomena at the global network scale, such as key properties of information diffusion~\cite{DBLP:journals/osnm/ArnaboldiCPD17}. Therefore, studying how independent avatars may modify, in the Metaverse, understanding and modelling structural properties of users social networks at the microscopic level, is a pre-requisite to study larger-scale effects at the global level of future Metaverse Online Social Networks. This methodology follows in the footsteps of prior work related to current Online Social Networks, i.e., modelling ego network structures first ~\cite{PASSARELLA20122201}, model global social networks as a coherent collection of individual ego networks next~\cite{Conti2012}, and study emerging phenomena at the overall network level, last~\cite{DBLP:journals/osnm/ArnaboldiCPD17}. This paper marks the first step in this journey. Specifically, we investigate how the Metaverse could modify in non-trivial ways ego networks, which is the required starting point to study the properties of large-scale Metaverse Online Social Networks.

By endowing avatars with autonomous capabilities and independence, users can potentially gain the ability to delegate (part of) their socializing responsibilities to their digital counterparts. This delegation enables users to maintain existing social contacts seamlessly, freeing up valuable non-avatar-mediated socializing time. Such liberated time could be strategically invested in additional socialization activities, expanding the user's virtual social footprint and fostering a richer and more diverse social experience. As it is known that the structural properties of users' social networks are, by and large, determined by the time individuals can allocate to socialising activities, independent avatars may enable users to augment their effective socialisation time, with possible profound impact on their social network structure~\cite{aral2023exactly}. As is the case of current online social network platforms, a detailed understanding of such phenomena can be the basis for new Social Network services in the Metaverse environment depicted in Figure~\ref{fig::logical}.
\emph{Specifically, in this work, we present an initial model of how an independent avatar could help its user to better navigate their social relationships and optimize their socializing time in the Metaverse by (partly) offloading some interactions to the avatar.}

Our computational modeling approach draws inspiration from quantitative models from social sciences, such as \emph{ego networks} \cite{DUNBAR201539}, \emph{social presence}  \cite{10.3389/frobt.2018.00114}, 
and \emph{social cueing} \cite{doi:10.1080/10447318.2023.2193514}, to build a microscopic characterization of social networking in the Metaverse era. 
By incorporating such concepts in our computational model, we aim to not only capture the complexities of users' social interactions but also provide a robust foundation for understanding how independent avatars can effectively navigate and augment the user social time capacity.

The roadmap of the paper is as follows. In Section \ref{sec:vision}, we delve into the vision and the challenges to be addressed. In Section \ref{sec:time_variables}, we define the setting and the variables that we are considering, by borrowing concepts from social science, human-computer interaction and online social networks fields. Specifically, we focus on defining the user's ego network, the avatar-mediated, non-avatar-mediated and debriefing time capacities, the social presence quantification, as well as the impact of social cues and compression. In Section \ref{sec::prob}, we define the problem of maximizing the non-avatar-mediated spare time and we explore its feasibility. Finally, in Section \ref{sec:results} we present some preliminary results about the optimization of the user spare time, which could be reinvested into the user social network.

\section{From microscopic modeling of social networks to large-scale Metaverse network properties} \label{sec:vision}

\subsection{Microscopic properties and large-scale social network properties}

Offline (commonly referred to as real-life) and online social networks have been subject to investigation for years. 
Typically, social networks can be studied from either a macroscopic or microscopic level. In the former, researchers analyze the overall structure of the network, encompassing factors such as size, density, centralization, clustering, and overall connectivity. This approach aims to comprehend the global properties and dynamics of the network and their broader influence on social phenomena. The macroscopic view can be characterized as a ``system'' perspective. Conversely, the microscopic view centers on individual actors (nodes) and their immediate connections and interactions within the network. The focus might be on individual node characteristics, such as centrality, influence, or similarity to others, as well as the dynamics of specific relationships, such as friendship, collaboration, or communication patterns \cite{10.5555/2901667}. 

However, microscopic structures do not solely exert microscopic effects; consider, for instance, ego networks. 
Ego networks serve as a graph-based framework commonly utilized to explore the social connections between an individual and their peers~\cite{everett2005ego,lin2001social,mccarty2002structure,hill2003social}. This abstraction holds significance as many aspects of social behavior, such as resource sharing, collaboration, and information diffusion, are primarily influenced by its structural characteristics~\cite{sutcliffe2012relationships}.  Prior work has shown that it is possible to build realistic large-scale social networks starting from ego network models capturing the characteristics of the relationships between egos and their alters. Specifically,~\cite{Conti2012} has defined a model for creating large scale networks connecting individual egos (and their alters) in such a way that microscopic properties are preserved~\cite{PASSARELLA20122201}. These networks created according to these model accurately reproduce key properties of large-scale social networks, such as short paths, degree distribution, clustering. Quite interestingly, these models have been used to study emerging properties at the macroscopic level, for instance in the case of information diffusion. It has specifically been shown that the well-known property of ``six degrees of separation” does not hold anymore in case of diffusion of \emph{trusted} information requiring certain level of trust among nodes in order to be propagated. In this case, it has been found that length of paths can be way longer from the length of the \emph{topological} shortest paths~\cite{DBLP:journals/comcom/ArnaboldiCGPP16, DBLP:journals/osnm/ArnaboldiCPD17, DBLP:journals/osnm/ArnaboldiCPD17}.

Now that Metaverse is emerging, possibly bearing a profound impact on the way we architect online social networks nowadays, we believe that reconsidering ego networks in the Metaverse, and the resulting properties of large-scale Metaverse Online Social Networks generated by them, is of paramount importance.

\subsection{Ego networks in the Metaverse}

\begin{figure}[t!]
\begin{center}
\includegraphics[width=0.5\columnwidth]{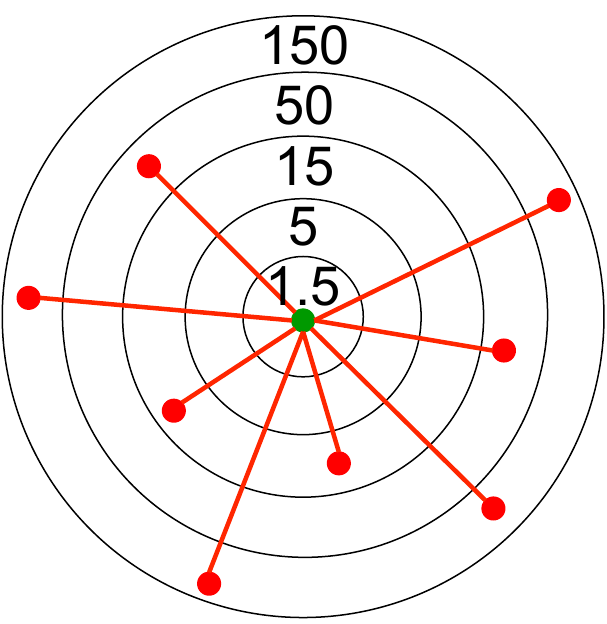}
\caption{Layered structure of human ego networks. The green node represents the ego, the red nodes represent its alters. The numbers give the size of the corresponding social circle.}
\label{fig:egonet}
\end{center}
\end{figure}

Within an ego network, the individual, known as the "ego," occupies the central position in the graph, while the edges represent connections to peers, referred to as "alters," with whom the ego interacts. The strength of the ego-alter ties is often quantified based on the frequency of interactions between them. By categorizing these ties according to their strength, a layered structure emerges within the ego network (see Figure~\ref{fig:egonet}), wherein the inner circles encompass the closest social connections, while the outer circles represent more distant relationships.

According to the \emph{social brain hypothesis} proposed in evolutionary psychology~\cite{dunbar1998social}, the existence and sizes of these groups are dictated by the maximum cognitive capacity of the human brain allocated to maintaining \emph{meaningful} social relationships—beyond mere acquaintances.  This model delineates five layers within the confines of the Dunbar number, which denotes the maximum number of social relationships (averaging around 150) that an individual can actively uphold~\cite{hill2003social, Zhou2005}. Relationships beyond this threshold are considered acquaintances, with their maintenance having minimal impact on cognitive resources. Figure~\ref{fig:egonet} depicts the typical sizes of each layer (1.5, 5, 15, 50, 150), with an approximate ratio of~3 between adjacent layers—a common observation in human ego networks~\cite{Zhou2005}.

The layered structure of ego networks serves as a hallmark of human social interactions, both online and offline. Given that the Metaverse represents a more immersive fusion of offline and online realms, it is anticipated that such a structure will largely persist, similarly to when social relationships split between the offline world and online social networks~\cite{DUNBAR201539,DBLP:conf/infocom/ArnaboldiCPP13}. However, a Metaverse augmented with independent avatars introduces a novel avenue for social interactions: human-to-avatar engagement. Unlike current cyber-physical systems where there are no avatar-mediated user-to-user interactions, the vision of an augmented Metaverse with independent avatars enables a new form of socialization, wherein a user interacts with an avatar acting on behalf of another user (who can engage in a different activity), or where two avatars interact between each other on behalf of their respective users. This presents an unprecedented opportunity to reshape social interactions. In essence, this quantitative model is dictated by the total time ego (as a human being) can allocate to socialisation, as a proxy of the cognitive effort entailed in maintaining active social relationships. Quite interestingly, nowadays technologies, including online social networks, cannot change the essence of this model, as none of them can increase the cognitive capacity human beings can allocate to socialisation~\cite{DUNBAR201539}. Therefore, the very same structure depicted in Figure~\ref{fig:egonet} has been observed in online social networks. 

In this context, our focus lies on the reallocation of socialization time from user-to-user (u2u) interactions to user-to-avatar (u2a) interactions. As previously mentioned, the presence of independent avatars frees up cognitive social capacity for users, allowing them to reinvest it in leisure activities or other social interactions. This newly available cognitive social capacity and its redistribution have the potential to reshape human social networks to some extent. For instance, users may choose to reinvest their freed-up social time into forming new weak ties within the network. Given that weak ties effectively bridge different topological groups in social graphs, this could lead to more efficient dissemination of information~\cite{granovetter1973strength}.

Therefore, we argue that understanding the interplay between ego network structures and the envisioned augmented Metaverse is essential for comprehending future Metaverse-based social life. In this study, we address the foundational aspect: how to (partly) offload the ego's socialization time to avatars in an ego network, chosen as a significant example of microscopic structures. 


\section{The setting and its variables} \label{sec:time_variables}

In this section, we provide a brief introduction to the ego network modelling and  we define the fundamental variables that express the various notions of time that we consider in our formulation. 

\subsection{User's ego network}

Following the Dunbar's model of ego networks (Figure~\ref{fig:egonet}, we assume that a user $u$ (the ego, in Dunbar's terminology) maintains a personal network that consists of the people $v_1, v_2, ..., v_n \in V$ (called alters) with whom $u$ interacts with regularly. 
Within this ego network, $u$ has different levels of closeness to each person. At the core of the network are a few people with whom $u$ has the closest relationships, while at the periphery there are acquaintances that $u$ only interacts with occasionally.
In an ego network, relationships are organized, according to their strength, in concentric circles (or layers). Typically, the ego network of $u$ consists of 4-5 layers.

It is important to note that $u$ may face limitations in terms of time and resources for socializing with other individuals. A major challenge is that $u$ only has a limited amount of time (and cognitive capacity). 
This may mean that $u$ must prioritize certain connections over others, and may not be able to invest in as many new relationships as they would like. Another challenge is that expanding the network may require the ego to navigate unfamiliar social contexts or communities, which can be intimidating or uncomfortable. It may take time and effort to build trust and establish rapport with individuals who are not part of the ego's existing social circles. Finally, there is also the risk of spreading oneself too thin by trying to maintain connections with too many individuals. 


\subsection{Socializing time capacities $\tilde{X}$, $Y$, debriefing time $Z$}

We assume user $u$ with corresponding independent avatar $a$ which can operate in the Metaverse also independently (when $u$ is offline). $u$ spends on average $\tilde{X}$ time (non-avatar-mediated) socializing. $\tilde{X}$ is partitioned across $u$'s contacts $v \in V$ in the ego network in a non-uniform manner, with 
\begin{equation} \label{eqn:Tu}
\tilde{X} = \tilde{x}_1 + \tilde{x}_1 + .... + \tilde{x}_n.
\end{equation}
A person spends on average  about 20\% of their time for socialization purposes \cite{dunbar1998theory}, so, for a given 8760 hours in a year, the expected value of non-avatar-mediated socializing time would be
\begin{equation}
\mathbf{E}[\tilde{X}] = 8760 \cdot 0.2 = 1752.
\end{equation}

Given that avatar $a$, when operating independently in the Metaverse, can contribute a fixed amount of socializing time $Y$ (avatar-mediated) to the current activities of $u$, we assume that $a$ can virtually augment $u$'s socializing time and therefore
enhance $u$'s socializing time across the existing contacts in the ego network,
partitioning its time in a non-uniform manner, with\footnote{It is therefore apparent that during the activity delegation from $u$ to $a$, $u$ might decide to reduce the amount of non-avatar-mediated time for contact $v$ and replace it with an amount of avatar-mediated time instead.}
\begin{equation}\label{eqn:Ta}
Y = y_1 + y_2 + .... + y_n.
\end{equation}
Indicatively, we can assume for example that the avatar can spend all of its available time socializing, so, for a given 8760 hours in a year, the expected value of non-avatar-mediated socializing time would be
\begin{equation}
\mathbf{E}[Y] = 8760.
\end{equation}

In order for $u$ to stay up-to-date with $a$'s activities and interactions, $u$ will need to sync with $a$ periodically to receive a debrief of its socialization updates for each $v$. This synchronization is crucial to maintain $u$'s cognitive engagement in these social relationships. According to the ego network model, without this cognitive involvement, relationships may transition out of the ego network and become mere acquaintances, as $u$ would not spend \emph{any} cognitive resources on those relationships, which would inevitably fade out with time. 
Thus, despite the potential resource costs, the debriefing process is necessary for staying informed and engaged with the $a$'s activities, as well as effectively incorporating the social presence value achieved by $a$. In order for $u$ to get debriefed by $a$ for all contacts in $V$, $u$ would need to spend 
    \begin{equation}
    Z = z_1 + z_2 + ... + z_n \label{eqn:Td}
\end{equation}
time. However, due to the fact that attention span (defined as the amount of time spent concentrating on a task before becoming distracted \cite{berger2018present}) can decrease in some given amount of time, leading to distractibility, we assume that the attention during the debriefing process is eventually uncontrollably diverted to another activity. We therefore impose an upper threshold of $Z_{\text{max}}$ on the debriefing time, with 
\begin{equation}
  Z \leq Z_{\text{max}}.  \label{eq:zmax}
\end{equation}
If, for example, according to \cite{doi:10.1152/advan.00109.2016}, the daily threshold is 50 minutes, then, for a given year, we would have a debriefing time of
\begin{equation}
Z_{\text{max}} \approxeq 304 \textnormal{ hours}.
\end{equation}
This threshold incurs the potential unavailability of $u$ to effectively get debriefed $\forall v \in V$.


\subsection{Social presence}

In order to formulate the avatar socializing capacity $Y$, we need to better understand the social presence capacity of an independent avatar $a$ in the Metaverse. Social presence is broadly defined as the subjective experience of being present with a ``real'' user and having access to her thoughts and emotions. 
Different forms of networked communication systems offer different levels of social presence. 
Therefore, in order to achieve the same social presence value by using two different networked communication means, a user might need to spend different amounts of time. Social presence is often evaluated under different contexts (for example, user-user interaction, user-avatar interaction), and as such, some diversity in measures cannot be avoided. 
In the context of our paper, it is necessary to understand how different technological features of the Metaverse networked communication influence perceptions of social presence so as to be able to define a social presence function~$f_{s}$. Defining the function will in turn enable us to characterize the relation between the avatar-mediated communication time $y$ and the non-avatar-mediated communication time $\tilde{x}$.

In the social psychology literature, our notion of ``independent avatar'' would naturally fall in the category of ``agent''. Research works which have investigated the impact of the (perceived) agency on social presence typically assume the user as an actual person or a computerized character prior to the interaction. As the survey \cite{10.3389/frobt.2018.00114} nicely points out, the majority of the related literature converges to the conclusion that avatars generally elicit greater social presence value than agents, in the sense that users typically feel higher levels of social presence when a virtual counterpart was thought to be controlled by an actual person rather than by a computer program.

Digital media vary in their ability to transmit social cues and thereby facilitate social presence 
through technology-mediated interpersonal communication. Media that are high in social presence are likely better at facilitating social connectedness because they are closer to face-to-face communication compared to media that are lower in social presence. We naturally assume that non-avatar-mediated communication can ideally achieve a higher social presence value compared to avatar-mediated communication. Therefore the social presence value achieved with a unit of time spent in the first case ($\tilde{x}_v$) does not equal the value with a unit of time in the second case ($y_v$), and we need a mapping $f_{s}$ to convert the social presence of the avatar to the equivalent social presence of the user. In other words, in order for $u$ to have achieved the same amount of social presence value with a target contact $v$ when using an independent avatar $a$, $a$ would have needed to spend 
    \begin{equation}
    y_v = f_{s}(\tilde{x}_v) = \beta_v \cdot \tilde{x}_v
    \end{equation}
    time, with, according to \cite{2012appel}, 
    \begin{equation}    
    \mathbf{E}[\beta_v] = \frac{4.182}{3.236} = 1.29,
    \end{equation}
where $\tilde{x}_v$ would be the time that $u$ would need to spend socializing with $v$ instead. $4.182$ is the average social presence value measured from human-human interactions in~\cite{2012appel}, while $3.236$ is the average social presence value measured in the same experiment for human-computer interactions.

\subsection{Social cues and compression}

We now better formulate the debriefing process which requires both time and attention from $u$. Depending on the frequency and complexity of $a$'s social interactions, the debriefing process may take varying amounts of time for each $v \in V$ and may need to be prioritized alongside other activities. For each $v \in V$, we assume that the required amount of time for $u$ to get debriefed is not more than the actual interaction time between $a$ and $v$ and that $z_v$ is a function of $y_v$, therefore:
\begin{equation}
z_v = f_{d}(y_v) \leq y_v.
\end{equation} 
The individual debriefing efficiency can be quantified through compression efficiency 
and avatar anthropomorphic and social cue measurements 
which can be expressed as follows:
\begin{equation}
z_v = c \cdot \delta \cdot y_v = \gamma \cdot y_v,
\label{eq:debriefing_efficiency}
\end{equation}
%
%
with $c$ being the compression ratio, $\delta$ being the anthropomorphism extent and the social cuing efficiency that $a$ can achieve for debriefing. The compression ratio captures the fact that regular human interactions are interspersed with downtime in the conversation and are both partially inefficient (e.g., due to momentary misunderstandings) and redundant (because we tend to repeat concepts multiple times). Thus, a 1-hour conversation can often be summarised in a much shorter time. The compression ratio expresses this shrinking of time when summarising the salient points of a conversation. $\delta$ captures the efficiency of an avatar to convey a certain message depending on its level of anthropomorphism and its ability to reproduce social cues. 

According to known works on compression ratios, such as  \cite{10.1145/312624.312665} and\cite{ghalandari-etal-2022-efficient}, we can set 
\begin{equation}
\mathbb{E}[c] = 0.54.
\end{equation}
Also, 
social presence can be measured in several directions (verbal, visual, auditory, etc.). \cite{doi:10.1080/10447318.2023.2193514} provides a quantitative survey-based assessment that captures both the visual and verbal efficiency of chatbots, using a 7-point Likert scale. As an example, in the best case, highly anthropomorphic chatbots with a human-like conversation style reach a score of~6 out of~7. We use these results to assign a value to $\delta$. Specifically, in the above example, we say that 
\begin{equation}
\mathbb{E}[\delta] = 7/6.
\end{equation}
Note also that $\delta$ is fixed for every ego, as it simply depends on the social presence properties of the avatar associated with the ego. Circling back to Equation~\ref{eq:debriefing_efficiency}, this means that $\gamma = c \cdot \delta$ can be set to $0.63$.

\section{Maximizing the user spare time} \label{sec::prob}

In this section we formulate the problem of maximizing the non-avatar-mediated spare time and we explore its feasibility.

\subsection{Problem formulation}

\begin{problem}
For a user $u$ and independent avatar $a$, given a set of contacts $V$ and their respective initial non-avatar-mediated time partitioning $\tilde{x_1} + ... + \tilde{x_n} = \tilde{X}$, maximize the spare time of $u$ (i.e., $\tilde{X} - X - Z$, where $X$ denotes the non-avatar-mediated time for user $u$ when using an avatar to partially offload social interactions) by also using $a$'s avatar-mediated time for socialization in a resulting total combined time partitioning $(x_1 + y_1) + ... + (x_n + y_n) = X + Y$. 
\end{problem}

Following the problem definition and the discussion in the previous sections, in order for $u$'s 
time to be augmented, $u$ could delegate to $a$ a portion of the socializing activity time. In order to do so, $u$ would have to exchange some of the actual non-avatar-mediated time for implementing $a$'s debriefing process, with a final target to achieve an augmented social presence with 
$a$'s avatar-mediated time. In this case, $u$ would be able to have an ego network comprising the contacts in $V$ 
in a resulting amount of time across $u$ and $a$ combined. 
%
%
%
The problem can now be framed as follows:

\begin{align}
  \min\quad & \sum_v x_v + \gamma \cdot y_v  \label{opt:obj_fun}\\\
\text{s.t.\quad} & \nonumber \\
& x_v + \frac{1}{\beta_v} \cdot y_v \geq \tilde{x}_v \label{opt:c1} &   & \forall v  \\
& \sum_v x_v + \gamma \cdot y_v \leq \tilde{X} \label{opt:c2}&   & \\
& \sum_v y_v \leq Y \label{opt:c3}&   & \\
& \sum_v \gamma \cdot y_v \leq Z \label{opt:c4}&   & \\ 
& x_v, y_v \geq 0 \label{opt:c5} &   & \forall v
\end{align}

The objective function (Eq.~\ref{opt:obj_fun}) minimizes the sum of the non-avatar-mediated time and the debriefing time, thus maximizing the available non-avatar-mediated spare time on the user side. Constraints in Eq.~\ref{opt:c1} guarantee that the resulting effective non-avatar-mediated and avatar-mediated socializing time for each $v$ is not less than the initial non-avatar-mediated time, i.e., that the resulting relationships are not modified in terms of cognitive involvement with respect to the case when the avatar is not present. Constraint in Eq.~\ref{opt:c2} guarantees that the non-avatar-mediated time and the debriefing time do not surpass the total available time of the user. Constraint in Eq.~\ref{opt:c3} guarantees that the avatar-mediated time does not surpass the total available time of the avatar. Constraint \ref{opt:c4} guarantees that the debriefing process abides by the non-distractibility rule introduced with Eq.~\ref{eq:zmax}. Finally, constraints in Eq.~\ref{opt:c4} guarantee that time values are zero or positive.

\subsection{Problem feasibility}

\newcommand{\cc}[1]{\textcircled{\raisebox{-0.9pt}{#1}}}

The problem we have formulated in the previous section can be solved using any standard LP solver. For the purpose of our investigation, it is interesting to analyse the feasible region of the problem, where a solution can be obtained. We assume here that $\beta$ and $\gamma$ are constant and equal for all alters. 

Due to the minimization nature of the problem, the first constraint (Eq.~\ref{opt:c1}) can only hold as an equality. Consequently, the constraint can be used to perform a variable substitution in the objective function, as $x_v$ can be expressed as a function of $y_v$ and the initial physical time allocated to that alter $\tilde{x}_v$. Specifically, it holds the following:
\begin{equation}
    x_v = \tilde{x}_v - \frac{1}{\beta} y_v.
    \label{eq:xv_sub}
\end{equation}
Effectively, we just need to find a solution for $y_v$ and the values for $x_v$ will follow. This allows us to rewrite the optimization problem as shown below. 

\begin{align} 
  \min\quad & \tilde{X} + \left(\gamma - \frac{1}{\beta}\right) \sum_v y_v \label{opt2:obj_fun} \\
\text{s.t.\quad} & \nonumber \\
& \left(\gamma - \frac{1}{\beta}\right) \sum_v y_v \leq 0 \label{opt2:c2}&   & \\
& \sum_v y_v \leq \min \left\{ Y, \frac{Z}{\gamma} \right\} \label{opt2:c3}&   & \\
& 0 \leq y_v \leq \beta \tilde{x}_v \label{opt2:c1} &   & \forall v
\end{align}


Let us now look at the first constraint (Eq.~\ref{opt2:c2}), which effectively requires that the debriefing time is always smaller than the time user $u$ should have spent directly interacting with contact $v$ (if this were not the case, user $u$ would be better off performing non-avatar-mediated interactions, as avatars would bring no benefit). Since, according to the constraint in Eq.~\ref{opt2:c1}, $y_v$ are always positive, the first constraint can only be satisfied if $\gamma \leq \frac{1}{\beta}$ (case \cc{A}) or if $y_v = 0$ for all $v$ (case \cc{B}). The discussion in Section~\ref{sec:time_variables} tells us that in a realistic setting $\frac{1}{\beta} \sim 0.78$ and $\gamma \sim 0.63$, so we are operating in case \cc{A} and non-trivial solutions (i.e., $y_v > 0$) are possible.
The remaining constraint (Eq.~\ref{opt2:c3}) is a simple cap on the total avatar time. Its concrete meaning is that the total time the avatar spends socializing cannot be greater than the overall time in the avatar's capacity for socialization or than the maximum debriefing capacity of the user.

\begin{figure}
    \centering
    \includegraphics[width=\linewidth]{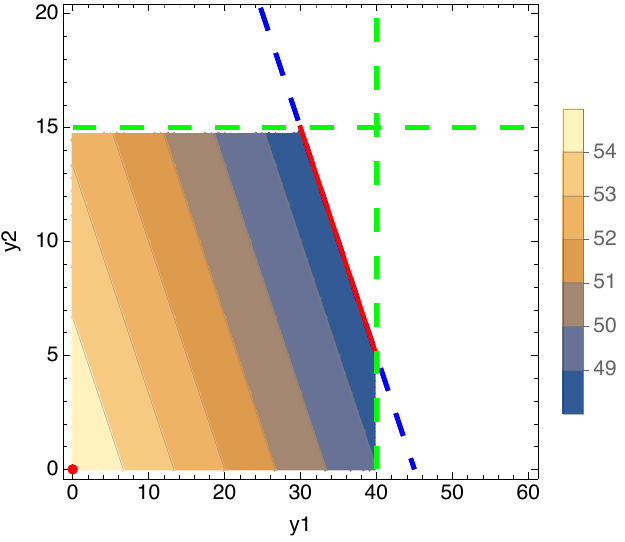} 
    \caption{Contour plot of the objective function in Eq.~\ref{opt2:obj_fun} when the socialization involves two alters. The dashed blue line corresponds to the constraint in Eq.~\ref{opt2:c3}, the dashed green lines to those in Eq.~\ref{opt2:c1}. The red line is the minimizing solution for case \cc{A}, the red point for case \cc{B}. The plot is obtained with $\tilde{x}_1 = 40, \tilde{x}_2 = 15, \gamma = 0.63, \frac{1}{\beta} = 0.78, Y = 45, Z = 45$.}
    \label{fig:feasible-region}
\end{figure}

In light of the reformulated problem, we can now revisit the objective function $\min \tilde{X} + (\gamma  - \frac{1}{\beta}) \sum_v y_v$. Let us distinguish the two cases, \cc{A} and \cc{B}, that we have already introduced. When $(\gamma  - \frac{1}{\beta}) > 0$ \cc{B}, the objective function is minimized when $y_v = 0$ for all $v$. Vice versa, when $(\gamma  - \frac{1}{\beta}) \leq 0$ \cc{A}, as in the realistic conditions discussed in Section~\ref{sec:time_variables}, the objective function is minimized when $y_v$ take their highest values within the feasible region. As illustrated in Figure~\ref{fig:feasible-region}, in case \cc{A}, the objective function is minimized when $y_v$ lies on the red solid line, while in case \cc{B} when $y_v = 0$. 

The above discussion allows us to identify two main regimes in our avatar-mediated socialization problem. On the one hand, when the debriefing process is less efficient than the avatar socialization process (i.e., when $\gamma > \frac{1}{\beta}$), delegating socialization to the avatar is not convenient because the time saved from socialization is spent in debriefing. The optimal solution in this case is $y_v = 0$, which is equivalent to not using the avatar at all. Vice versa, when the debriefing process is more efficient than the avatar socialization process (i.e., when $\gamma \leq \frac{1}{\beta}$), the optimal solution consists in using the avatar as much as possible, or at least as much as the constraints allow. In short, with these linear constraints and objective function, depending on the socialization and debriefing efficiencies, the user should either not use the avatar at all or delegate as much as possible to it. This conclusion aligns intuitively with our problem formulation.

\section{Insights from the model}
\label{sec:results}

\begin{figure}
    \centering
    \includegraphics[width=\linewidth]{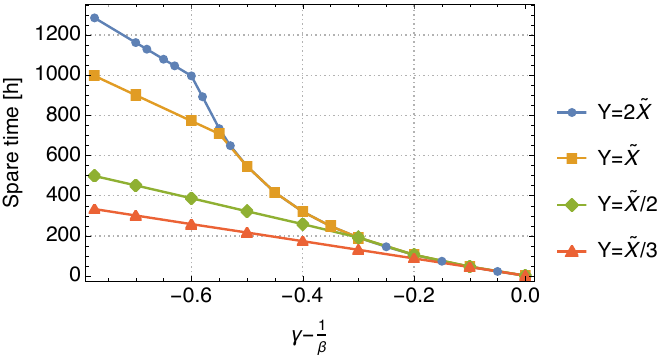}
    \caption{Spare time for an ego network with $|V| = 117, \tilde{X} = 1288h, Z = 300h, \beta = 1.29$.}
    \label{fig:spare-time}
\end{figure}

We wrapped the previous section with a qualitative finding: when the debriefing process is more efficient than the avatar socialization process, using the avatar is always convenient. In this section, we delve into a quantitative analysis of this benefit. To this aim, we leverage the model provided in~\cite{Conti2012} for generating realistic ego networks. We fix $\beta$ to $1.29$ in our optimization problem and we vary $\gamma$ between $[0,\frac{1}{\beta})$ (i.e., we remain in the range where case \cc{A} holds but we assume, as it is realistic, that there will be some variability in the $\gamma$ values). We test scenarios where the total avatar time $Y$ is greater, equal, or smaller than the total user socialization time without the avatar ($\tilde{X}$). In Figure~\ref{fig:spare-time}, we plot the amount of spare time gained by the user by leveraging the avatar. When $\gamma$ and $\frac{1}{\beta}$ are close (i.e., their difference is near zero), the gained spare time is very small. However, as the debriefing becomes more and more efficient with respect to the avatar socialization, then the time saved by the user becomes increasingly higher. 

To quantify the potential impact of the spare time, we compare the saved time against the time an average user spends on individual alters of its ego network layers. We extract these numbers from~\cite{Conti2012}, where statistics for three concentric layers are reported. The layers are referred to as the ``support clique'', ``sympathy group'', and ``active network'', with average cumulative sizes of 4.6, 14.3, and 132.5 alters, respectively. This structure deviates from the standard ego network structure due to the exclusion of the ``affinity group" layer, whose properties are not well-defined in the related literature.
From~\cite{Conti2012}, we obtain $t_{\textrm{active}}=8.81 h$, $t_{\textrm{sympathy}} = 38.72 h$, $t_{\textrm{support}}=74 h$.
Considering together the time spent on all the alters in each layer, we obtain $T_{\textrm{active}}=1041.51 h$, $T_{\textrm{sympathy}} = 375.83 h$, $T_{\textrm{support}}=340.65 h$. 
%
It is evident that when the amount of time saved is high enough, users might have the opportunity to reinvest it in expanding their ego network circles and enhancing their social connections. For instance, with approximately 1000 spare hours, one could potentially double the active layer of their ego network, or with around 350 spare hours, strengthen their support clique. While fully characterizing the dynamics of this augmentation is beyond the scope of this work, and the relationship between time saved and new connections gained might not be as direct and straightforward, these initial insights suggest that avatars in the Metaverse have the potential to increase a user's available time -- a resource that could be redirected toward growing their personal social network. In theory, this could result in surpassing Dunbar's number of 150 meaningful relationships per ego, thus fostering a significantly enriched social life within the Metaverse -- an advancement not observed in online relationships within the social web~\cite{DUNBAR201539}.

\section{Conclusion and discussion}
\label{sec:conclusion}

This work's primary contribution lies in addressing structural properties of human social networks in the Metaverse, through an approach rooted in social sciences findings. We posit that this is a fundamental building block for the design of Metaverse social networks, as similar models have been for the design of online social networks. Leveraging those concepts, our approach is as follows: 
\begin{enumerate}
    \item Navigate the network of personal connections within the Metaverse through the anthropological lens of ego networks, focusing on the user related parameter configuration and the relation between physical and digital interactions. 
    \item Incorporate social presence concepts to allows us to delve into the perceptual aspects of these digital relationships, emphasizing the sense of being together in the virtual space and the impact of this presence on users' social experiences. 
    \item Integrate insights from social cueing observations, so as to quantify cues embedded in digital interactions to enhance the authenticity and richness of social experience modelling. 
\end{enumerate}
The paper introduces an innovative model that delineates user-mediated and avatar-mediated interactions, synthesizing them into a coherent social cognitive involvement framework. Our focus has been on enhancing user socialization by streamlining interactions through avatar delegation, serving as a foundational step for a more comprehensive model of the characteristics of users' social networks in the Metaverse. 

Immediate advancements could involve exploring various dependencies among problem variables, surpassing mere linearity. Expanding the scope, one intriguing direction involves integrating the concept of social circles present in Dunbar's ego network models, an aspect overlooked in our current paper. Additionally, a promising research direction involves determining the minimum level of user-mediated interactions necessary to maintain relationships within the ego network. Building upon this, we aim to explore how users might reinvest the time saved by avatars into forging new social connections and whether the assistance of avatars leads to a restructuring of user ego networks.

\section*{Acknowledgments}
This work was partially supported by SoBigData.it, which receives funding from the European Union -- NextGenerationEU -- National Recovery and Resilience Plan (Piano Nazionale di Ripresa e Resilienza, PNRR) – Project: ``SoBigData.it – Strengthening the Italian RI for Social Mining and Big Data Analytics'' -- Prot. IR0000013 -- Avviso n. 3264 del 28/12/2021. Chiara Boldrini was supported by PNRR - M4C2 - Investimento 1.4, Centro Nazionale CN00000013 - ``ICSC -- National Centre for HPC, Big Data and Quantum Computing'' -- Spoke 6, funded by the European Commission under the NextGeneration EU programme. Marco Conti and Andrea Passarella were partly supported by PNRR -- M4C2 -- Investimento 1.3, Partenariato Esteso PE00000013 -- ``FAIR - Future Artificial Intelligence Research'' -- Spoke 1 ``Human-centered AI'', funded by the European Commission under the NextGeneration EU programme.

\balance
\bibliographystyle{IEEEtran}
\bibliography{references}

\end{document}